\documentclass[12pt]{article}
\usepackage[a4paper, hmargin=2.5cm, vmargin=2.5cm]{geometry}
\usepackage{amsfonts,amsmath}
\usepackage[authoryear]{natbib} \bibpunct{(}{)}{,}{a}{,}{,}

\newcommand{\bA}{\mbox{\boldmath $A$}}
\newcommand{\bY}{\mbox{\boldmath $Y$}}
\newcommand{\bU}{\mbox{\boldmath $U$}}
\newcommand{\bC}{\mbox{\boldmath $C$}}
\newcommand{\bV}{\mbox{\boldmath $V$}}
\newcommand{\bI}{\mbox{\boldmath $I$}}
\newcommand{\bX}{\mbox{\boldmath $X$}}
\newcommand{\bB}{\mbox{\boldmath $B$}}
\newcommand{\bD}{\mbox{\boldmath $D$}}
\newcommand{\ba}{\mbox{\boldmath $a$}}
\newcommand{\by}{\mbox{\boldmath $y$}}
\newcommand{\bd}{\mbox{\boldmath $d$}}
\newcommand{\be}{\mbox{\boldmath $e$}}
\newcommand{\bq}{\mbox{\boldmath $q$}}
\newcommand{\bu}{\mbox{\boldmath $u$}}
\newcommand{\bx}{\mbox{\boldmath $x$}}
\newcommand{\bpi}{\mbox{\boldmath $\pi$}}
\newcommand{\bPsi}{\mbox{\boldmath $\Psi$}}
\newcommand{\btheta}{\mbox{\boldmath $\theta$}}
\newcommand{\bepsilon}{\mbox{\boldmath $\epsilon$}}
\newcommand{\bmu}{\mbox{\boldmath $\mu$}}
\newcommand{\bSigma}{\mbox{\boldmath $\Sigma$}}
\newcommand{\bOmega}{\mbox{\boldmath $\Omega$}}
\newcommand{\bdelta}{\mbox{\boldmath $\delta$}}
\newcommand{\bDelta}{\mbox{\boldmath $\Delta$}}
\newcommand{\bLambda}{\mbox{\boldmath $\Lambda$}}
\newcommand{\bbeta}{\mbox{\boldmath $\beta$}}
\newcommand{\bzero}{\mbox{\boldmath $0$}}
\newcommand{\bOne}{\mbox{\boldmath $1$}}

\begin{document}
\title{On formulations of skew factor models: skew errors versus skew factors}
\author{Sharon X. Lee$^1$, Geoffrey J. McLachlan$^{1, \star}$}
\date{}

\maketitle

\begin{flushleft}
$^1$Department of Mathematics,
University of Queensland, St. Lucia, Queensland, 4072, Australia.\\
$^\star$ E-mail: g.mclachlan@uq.edu.au 
\end{flushleft}

\begin{abstract}
In the past few years, there have been a number of proposals for generalizing the factor analysis (FA) model and its mixture version (known as mixtures of factor analyzers (MFA)) using non-normal and asymmetric distributions. These models adopt various types of skew densities for either the factors or the errors. While the relationships between various choices of skew distributions have been discussed in the literature, the differences between placing the assumption of skewness on the factors or on the errors have not been closely studied.    
This paper examines these formulations and discusses the connections between these two types of formulations for skew factor models.  
In doing so, we introduce a further formulation that unifies these two formulations; that is, placing a skew distribution on both the factors and the errors.      
\end{abstract}

\section{\large Introduction}
\label{s:intro}

Mixture models with skew component densities have gained increasing attention in recent years due to their ability in accommodating asymmetric distributional features in the data. 
However, these models are highly parametrized and so are not well suited for the analysis of high-dimensional datasets. 
One approach to reduce the number of unknown parameters in these models is to adopt a mixture of factor analzyers (MFA) model \citep{J617,J708,B004}.  
Recent developments along this path have explored factor-analytic equivalent of these skew mixture models for modelling high-dimensional datasets. To name a few, there are mixtures of (generalized hyperbolic) skew $t$-factor analyzers (GHSTFA) by \citet{J618}, the skew $t$-factor analysis (STFA) model by \citet{J160b}, the mixtures of generalized hyperbolic factor analyzers (GHFA) by \citet{J620}, the mixtures of skew normal factor analyzers (MSNFA) by \citet{J159b}, and more recently, the mixtures of hidden truncation hyperbolic factor analyzers (HTHFA) and scale mixtures of canonical fundamental skew normal factor analyzers (SMCFUSNFA) by \citet{J615} and \citet{J638}, respectively.  
\newline 

There are distinct differences between these factor-analytic models
available in the literature, not only on the choice of component densities, but also on
where the assumption of skewness is placed in the model (that is, whether
it is assumed for the factors and/or for the errors). 
The former had been considered by \citet{J638} and \citet{J105,J103}, who provide an account of existing models and discuss the links and relationships between the different component densities adopted by these models. 
Here we consider the implications of placing a skew distribution on the factors, or on the errors, or both. 
It should be noted that, to our knowledge, in all of the existing models, the assumption of skewness is placed  \emph{either} on the factors or the errors, but not both. 
A summary of these models is given in Tables \ref{tab:SE} and \ref{tab:SF} for models with skew errors (SE) and skew factors (SF), respectively.  
In order to study the differences between them, we consider yet another model that is more general - a factor analysis model with skew distributions for both the factors and the errors, namely, a SFE model.     
\newline

In this paper, we study the SE, SF, and SFE models and discuss their properties.  We provide a brief background on FA and skew models in Section \ref{s:background}, including summaries in tables listing major existing SE and SF models. 
The SFE model is introduced in Section \ref{section:SFE}. 
This model and the nested SE and SF models can be fitted by maximum likelihood via an expectation--maximization (EM) algorithm \citep{J034}; more specifically, an alternating expectation conditional maximization (AECM) algorithm \citep{J631} is used. These algorithms are derived in Section \ref{sec:EM}. 
   
\begin{table}
	\centering
	\hspace*{-1cm} 
		\begin{tabular}{|c||c|c|c|c|}
			\hline
			\textbf{SE Models} & \textbf{Notation} & \textbf{Factors} 
			& \textbf{Errors} & \textbf{References} \\
			\hline \hline
			Generalized hyperbolic 	&	MGHFA &	SGH 	& GH & \citet{J620} \\\hline  
			Generalized hyperbolic skew $t$ 		&	MGHSTFA &	$t$ & GHST & \citet{J618} \\\hline 
			CFUSN 	&	CFUSNFA &	normal 	& CFUSN & \citet{J624} \\\hline  
			Unrestricted skew $t$ 	&	uMSTFA & $t$ 		& uMST & \citet{J623} \\\hline 
		\end{tabular}
	\caption{Factor analysis (FA) and Mixtures of factor analyzers (MFA) models 
					with skew errors. 
					The notation GH, GHST, CFUSN, and uMST denote the generalized hyperbolic 
					distribution, the generalized hyperbolic skew $t$-distribution, 
					the canonical fundamental skew normal distribution, 
					the unrestricted multivariate skew $t$-distribution, 
					and the variance gamma distribution, respectively. 
					The prefix `S' in SGH denotes the symmetric version of the GH distribution.   }
	\label{tab:SE}
\end{table}

\begin{table}
	\centering
	\hspace*{-1cm} 
		\begin{tabular}{|c||c|c|c|c|}
			\hline
			\textbf{SF Models} & \textbf{Notation} & \textbf{Factors} 
			& \textbf{Errors} & \textbf{References} \\
			\hline \hline
			Restricted skew normal 	&	MSNFA &	rMSN & normal & \citet{J159b} \\\hline  
			CFUSH$^*$  &	CFUSHFA 	&	CFUSH & SGH & \citet{J615} \\\hline  
			Restricted skew $t$ 		&	MSTFA &	rMST & $t$ & \citet{J160b,J612} \\\hline 
			SMCFUSN					&	SMCFUSNFA 	&	SMCFUSN & SMN & \citet{J638} \\\hline
			CFUSN						&	CFUSNFA 	&	CFUSN & normal & \citet{J638} \\\hline   
			CFUST 					&	CFUSTFA 	&	CFUST & $t$ & \citet{J638} \\\hline 
		\end{tabular}
	\caption{FA and MFA models with skew factors. 
					The notation rMSN, CFUSH, SMN, and CFUST denote the 
					restricted multivariate skew normal distribution, 
					the canonical fundamental skew (symmetric generalized) hyperbolic distribution,
					a scale mixture of normal distributions, 
					and the canonical fundamental skew $t$-distribution, respectively. 
					$^*$The CFUSH distribution is not identifiable 
					and hence \citet{J615} imposed constraints on the parameters 
					to achieve identifiability and called it 
					the hidden truncation hyperbolic (HTH) distribution.}
	\label{tab:SF}
\end{table}

\section{\large Background}
\label{s:background}
\noindent

Skew distributions adopted in the above-mentioned models have a stochastic representation in the form of the convolution of a symmetric random variable and a `skewing' variable. For example, the canonical fundamental skew normal (CFUSN) distribution has a convolution-type stochastic representation given by the sum of a half normal and a normal variate. More formally, let $\bY$ be a $p$-dimensional random vector. If $\bY$ follows a CFUSN distribution, it can be expressed as
\begin{eqnarray}
\bY = \bmu + \bDelta |\bU| + \bV, 
\label{CFUSN}
\end{eqnarray}
where $\bU \sim N_q(\bzero, \bI_q)$ independently of $\bV \sim N_p(\bzero, \bSigma)$.
The parameter $\bmu$ is a $p$-dimensional vector of location parameters and $\bDelta$ is a $p\times r$ matrix of skewness parameters. 
We write $\bY \sim \mbox{CFUSN}_{q,r}(\bmu, \bSigma, \bDelta)$ if $\bY$ is generated from (\ref{CFUSN}). To simplify the discussion, we shall refer to $\bU$ as the \emph{skewing} variable and $\bV$ as the \emph{symmetric} variable. 
We note that all models listed in Table \ref{tab:SF} 
belong to the class of scale mixtures of CFUSN distributions (SMCFUSN). The latter has a stochastic representation similar to (\ref{CFUSN}), but with an additional (scalar) scaling variable $W$ on the covariance matrix of $\bU$ and $\bV$; that is, it is given by 
$\bY = \bmu + \sqrt{W} (\bDelta |\bU| + \bV)$.  
On the other hand, the MGHSTFA model is a limiting case of the MGHFA model, 
which is a variance-mean mixture of the normal distribution given by $\bY = \bmu + W \bdelta  + \sqrt{W} \bV$.  
To simplify the discussion, we will use the CFUSN distribution as an illustration,  
but note that analogous arguments apply to the SMCFUSN and GH distributions.  
\newline

\noindent
The traditional factor-analytic (FA) approach (applied to a random vector $\bY \in \mathbb{R}^p$ that has a normal distribution) is to decompose $\bY$ into a lower-dimension vector of factors $\bX$ and a vector of errors $\be$ by letting 
\begin{eqnarray}
\bY = \bmu + \bB \bX  + \be, 
\label{FA}
\end{eqnarray}
where $\bX \sim N_q(\bzero, \bI_q)$ contains the latent \emph{factors} and $\be \sim N_p(\bzero, \bD)$ contains the \emph{errors}. The matrix $\bB$ is a $p\times q$ matrix of factor loadings and $\bD$ is a diagonal matrix ($\bD = \mbox{diag}(\bd)$ and $\bd \in\mathbb{R}^p$). 
The latter matrix $\bD$ is taken to be diagonal since it is assumed that the variables in $\bY$ are conditionally independent given $\bX$.  
Thus, the marginal distribution of $\bY$ is given by $\bY \sim N_p(\bmu, \bB\bB^T + \bD)$. 
In the case where $q>1$, the FA model (\ref{FA}) has an identifiability issue due to $\bB\bX$ being rotationally invariant, that is, the model is still satisfied if $\bX$ is pre-multiplied by an orthogonal matrix of order $q$ and $\bB$ is post-multiplied by the the transpose of the same matrix. A common approach is to impose $q(q-1)/2$ constraints on $\bB$ so that (\ref{FA}) can be uniquely defined. 
 
It is apparent from the above that the CFUSN model (\ref{CFUSN}) has the form as (\ref{FA}), by considering the $|\bU|$ as `factors' and the $\bU_1$ as `errors'. This implies the CFUSN model is a factor model with half-normal `factors' and normal `errors'. 
To avoid confusion, we shall refer to $\bX$ in (\ref{FA}) as \emph{factors} and $\be$ in (\ref{FA}) as \emph{errors}. 

It is clear from the above definitions that there can be different ways to generalize the FA model to a CFUSN factor analysis model, by combining (\ref{CFUSN}) and (\ref{FA}) in different ways. 
An immediate question is whether to incorporate the factor analytic form for the distribution of the skewing variables or for the symmetric variables, or even for both.
We will now consider each of these cases. 
\newline

\section{\large Three formulations of skew factor models}
\label{section:SFE}

\subsection{The skew errors (SE) model}
\label{sec:SE}

One of the more straightforward approaches is to decompose the symmetric latent variable (that is, the `error' term $\bV$ in (\ref{CFUSN})) into the factor-analytic form (\ref{FA}). Hence, the `factors' have a normal distribution (in the case of the CFUSN model), whereas the errors have a skew distribution. More specifically, we take $\bV = \bB\bX+\be$, so that $\bV\sim N_p(\bzero, \bB\bB^T + \bD)$. 
Thus, $\bSigma = \bB\bB^T + \bD$.  
  
Proceeding from (\ref{CFUSN}), we see that 
\begin{eqnarray}
\bY &=& \bmu + \bDelta |\bU| + \bV \nonumber\\
		&=& \bmu + \bDelta |\bU| + \left(\bB \bX  + \be\right)  \label{intermediate}\\
		&=& \bmu + \bB \bX + \bepsilon,
\label{error}
\end{eqnarray}
where now the `errors' $\bepsilon = \bDelta |\bU| + \be$ follow a $CFUSN_{p,r}(\bzero, \bD, \bDelta)$ distribution and the `factors' $\bX \sim N_p(\bzero, \bI_q)$ remain unchanged (from the normal factor model (\ref{FA})). 
It follows that the marginal density of $\bY$ is  
\begin{eqnarray}
\bY &\sim& CFUSN_{p,r}(\bmu, \bB\bB^T+\bD, \bDelta).
\label{error3}
\end{eqnarray}
Alternatively, we may also consider taking $\be$ in (\ref{FA}) to have a (central) \linebreak CFUSN distribution with stochastic representation given by (\ref{CFUSN}) to arrive at an equivalent expression to (\ref{error}). 

With this model, $\bY$, $\bX$, and $\be$ have expected value given by $\bmu+\sqrt{2/\pi}\bDelta \bOne_r$, $\bzero$, and $\sqrt{2/\pi}\bDelta\bOne_r$, respectively. Their corresponding covariance matrix is given by $\bB\bB^T+\bD + (1-2/\pi)\bDelta\bDelta^T$, $\bI_q$, and $\bD + (1-2/\pi)\bDelta\bDelta^T$, respectively.  

An advantage of the SE model is that it is relatively straightforward to construct and facilitates easy implementation of the AECM algorithm. 
In the mixture model case, the latter is essentially the simple combination of the EM implementation for the CFUSN mixture model and the MFA model, where the first cycle is identical to that for the EM algorithm for the CFUSN mixture model (except that $\bSigma$ is not estimated in the M-step) and the second cycle is identical to that for the MFA model.  

Existing SE models include the (unrestricted) skew $t$-MFA model \citep{J623}, 
the generalized hyperbolic skew $t$-MFA model \citep{J623}, 
and the specialized generalized hyperbolic MFA model \citep{J401}; 
see Table \ref{tab:SE}. 
With these models, the errors are assumed to follow the (unrestricted) skew $t$, generalized hyperbolic skew $t$, variance gamma, and hyperbolic distribution, respectively. The factors have the corresponding symmetric version of the distribution of the errors.

\subsection{The skew factors (SF) model}
\label{sec:SF}

Perhaps a more natural approach is to replace the factors in (\ref{FA}) with a CFUSN random variable. In this case, we let $\bX$ in (\ref{FA}) have the stochastic representation (\ref{CFUSN}). Note that we are only introducing skewness to $\bX$ and hence we take $\bX \sim CFUSN_{q, r}(\bzero, \bI_q, \bDelta)$, that is, $\bX$ has location parameter $\bzero$ and scale matrix $\bI_q$. Thus, $\bU \sim N_r(\bzero, \bI_r)$ and $\bV \sim N_q(\bzero, \bI_q)$. However, it is important to note that $E(\bX) \neq \bzero$ and $\mbox{cov}(\bX) \neq \bI_q$. The reader is referred to the properties of the CFUSN distribution \citep{J008}. 

It follows that
\begin{eqnarray}
\bY &=& \bmu + \bB \bX  + \be \label{factor} \\
		&=& \bmu + \bB \left(\bDelta |\bU| + \bV\right) + \be \nonumber\\
		&=& \bmu + \left(\bB \bDelta\right) |\bU| + \left(\bB \bV + \be\right) \label{middle} \\
		&=& \bmu + \bDelta^* |\bU| + \left(\bB \bV + \be\right),
\label{factor2}
\end{eqnarray}
where $\bDelta^* = \bB \bDelta$, the `factors' $\bX \sim CFUSN_{q,r}(\bzero, \bI_q, \bDelta)$, 
and the `errors' $\be \sim N_p(\bzero, \bD)$ 
remain unchanged (from the normal factor model (\ref{FA})).  
It follows that the marginal density of $\bY$ is 
\begin{eqnarray}
\bY &\sim& CFUSN_{p,r}(\bmu, \bB\bB^T+\bD, \bB\bDelta),
\label{factor3}
\end{eqnarray}
which is almost the same as the SE case (\ref{error3}). 
If we replace $\bDelta$ in (\ref{error3}) with $\bDelta^* = \bB \bDelta$, we obtain the SF model from the SE model. 
\newline

With this model, $\bY$, $\bX$, and $\be$ have expected value given by $\bmu+\sqrt{2/\pi}\bB\bDelta \bOne_r$, $\sqrt{2/\pi}\bDelta\bOne_r$, and $\bzero$, respectively. Their corresponding covariance matrix is given by 
$\bB\bB^T+\bD + (1-2/\pi)\bB\bDelta\bDelta^T\bB^T$, $\bI_q + (1-2/\pi)\bDelta\bDelta^T$, and $\bD$, respectively. 
Some authors choose to normalize the factor so that $E(\bX)=\bzero$ and $\mbox{cov}(\bX) = \bI_q$; see, for example, the MSNFA model \citep{J159b}, the MSTFA model \citep{J160b,J612}, and the CFUSSH model \citep{J615}. In this case, the distribution of $\bX$ needs to be appropriately reparametrized. It follows that the mean and covariance matrix of $\bY$ do not involve $\bDelta$ and are the same as that for the corresponding symmetric MFA model. In the case of a CFUSNFA model, for example, the vector of factors $\bX$ has the distribution $CFUSN_{q,r}(-\bA^{-\frac{1}{2}}\bDelta\bOne_r, \bA^{-1}, \bA^{-\frac{1}{2}}\bDelta)$, where $\bA = \bI_q + (1-2/\pi)\bDelta\bDelta^T$, and $E(\bY) = \bmu$ and $\mbox{cov}(\bY) = \bB\bB^T +\bD$.    
\newline
      			
Existing SF models include, for example, 
the (restricted) skew normal MFA model \citep{J159b}, 
the (restricted) skew $t$-MFA model \citep{J612}, 
the canonical fundamental skew hyperbolic MFA model \citep{J615},
and the canonical fundamental skew $t$-MFA model \citep{J638}; 
see Table \ref{tab:SF}. As noted in \citet{J638} 
the above-mentioned models belong to the class of scale mixtures of CFUSN factor analyzers.   
\newline

We can see from the above that the SE and SF models are very similar. 
Indeed, they seem to share an intermediate form given by (\ref{intermediate}) and (\ref{middle}). 
Consider the following intermediate representation that is the same as (\ref{intermediate}) above,
\begin{eqnarray}
\bY = \bmu + \bDelta_0 |\bU| + \bB \bV + \be,
\label{both}
\end{eqnarray} 
where $\bDelta_0$ is a $p \times r$ matrix  
and $\bU$, $\bB$, $\bV$, and $\be$ are as defined in (\ref{CFUSN}) and (\ref{FA}) above. 
If we take $\bV$ as the factors, we obtain the SE model.  
In the case of the SF model, we include the skewness term (that is, the second term on the right-hand side of (\ref{both})) as part of the factors and hence we write $\bDelta_0$ in terms of $\bB$ and $\bDelta$; that is, $\bDelta_0 = \bB\bDelta$. 
Hence, for both the SE and SF models, the unknown parameters are given by $\btheta = (\bmu, \bB, \bD, \bDelta)$, but $\bDelta$ is a $p\times r$ matrix in the SE case, whereas it is a $q\times r$ matrix in the SF case.  
Due to this, the SF model has a slightly lower number of free parameters than the SE model (assuming $q < p$).

\subsection{The skew factors and errors (SFE) model}
\label{sec:SFE}

The third and more involved approach is to allow both the factors $\bX$ and the errors $\be$ in (\ref{FA}) to have a CFUSN distribution, that is, combining the SE and SF approaches. In this case, we take $\bX \sim CFUSN_{q, r}(\bzero, \bI_q, \bDelta_0)$ as in the case of the SF model, but we also let $\be$ follow a CFUSN distribution with skewness matrix $\bDelta_1$, that is, $\be \sim CFUSN_{p, s}(\bzero, \bD, \bDelta_1)$. It is clear that this SFE model is a generalization of the SE and SF models, which can be obtained by taking $\bDelta_0 = \bzero$ and $\bDelta_1=\bzero$, respectively. 

It follows that the SFE model is given by
\begin{eqnarray}
\bY &=& \bmu + \bB \bX  + \be \\
		&=& \bmu + \bB \left(\bDelta_0 |\bU_0| + \bV_0\right) 
				+ \left(\bDelta_1 |\bU_1| + \bV_1\right) \nonumber\\
		&=& \bmu + \begin{bmatrix} \bB\bDelta_0 & \bDelta_1\end{bmatrix}
				\begin{bmatrix} |\bU_0| \\ |\bU_1| \end{bmatrix}
				+ \begin{bmatrix} \bB & \bI_p\end{bmatrix} 
				\begin{bmatrix} \bV_0 \\ \bV_1\end{bmatrix}.
\label{error_factor}
\end{eqnarray}
In this case, we have a linear combination of CFUSN distributions. 
Given that $\bU$ and $\bV$ are independent, the CFUSN distribution is closed under convolution. Hence, $\bY$ has a CFUSN distribution. This can also be seen from (\ref{error_factor}) above, where it can be deduced that
\begin{eqnarray}
\bY &\sim& CFUSN_{p,r+s}\left(\bmu, \bB\bB^T + \bD, \tilde{\bDelta} \right), 
\label{SFE}
\end{eqnarray}
where $\tilde{\bDelta} = \begin{bmatrix} \bB\bDelta_0 & \bDelta_1\end{bmatrix}$.   

With this model, $\bY$, $\bX$, and $\be$ have expected value given by \linebreak $\bmu+\sqrt{2/\pi}(\bB\bDelta_0 + \bDelta_1) \bOne_r$, $\sqrt{2/\pi}\bDelta_0\bOne_r$, and $\sqrt{2/\pi}\bDelta_1\bOne_s$, respectively. Their corresponding variance matrix is given, respectively, by 
\begin{eqnarray}
\mbox{cov}(\bY) &=& \bB\bB^T+\bD + \left(1-\frac{2}{\pi}\right) 
				\left(\bB\bDelta_0\bDelta_0^T\bB^T+\bDelta_1\bDelta_1^T\right), \nonumber\\
\mbox{cov}(\bX) &=& \bI_q + \left(1-\frac{2}{\pi}\right)\bDelta_0\bDelta_0^T, \nonumber\\
\mbox{cov}(\be) &=&  \bD + \left(1-\frac{2}{\pi}\right)\bDelta_1\bDelta_1^T. 
\end{eqnarray}
   
In the case of skew elliptical distributions, the requirement for closure under convolution is that $\bU$  and $\be$ are uncorrelated. Hence, a similar model can be constructed using these distributions. For example, in the case of a joint CFUST distribution for $\bU$ and $\be$ (not independent but uncorrelated), we have that $\bY$ also follows a CFUST distribution. 
In a similar way, a SFE model can be constructed using a CFUSH distribution.

\section{Parameter estimation via the ECM algorithm}
\label{sec:EM}

All three formulations of skew factor models described above can be fitted via an EM algorithm, namely, an AECM algorithm. We will consider the mixture model case for generality. 
In this case, the density of a $g$-component MFA model is given by
\begin{eqnarray}
f(\by; \bPsi) &=& \sum_{i=1}^g f(\by; \btheta_i), 
\label{eq:MFA}
\end{eqnarray}
where $f(\by; \btheta_i)$ denotes the density of the $i$th component of the mixture model with parameters $\btheta_i$ $(i=1, \ldots, g)$. The vector $\bPsi$ contains all unknown parameters of the mixture model. The $\pi_i$ $(i=1, \ldots, g)$ denote the mixing proportions, which are non-negative an sum to one. 

In the first cycle of the AECM algorithm, the missing data include the latent component labels $z_{ij}$ and latent skewing variable $\bU_{ij}$. The M-step in this cycle involves updating $\pi_i$, $\bmu_i$, $\nu_i$, and also $\bDelta_i$ (in the SE case only). In the second cycle of the AECM algorithm, the missing data include the $z_{ij}$ and latent factors $\bX_{ij}$. The parameters related to the latent factors which include $\bB_i$ and $\bD_i$ are updated on the M-step of this cycle. In the case of the SF model, $\bDelta_i$ is also updated in this cycle. 

For generality, we henceforth consider the case of a mixture of CFUST factor analyzers (CFUSTFA). The CFUSN factor analysis model described above is a limiting case of the CFUSTFA model as $\nu \rightarrow \infty$ and $g=1$ component. An outline of the AECM algorithm for the SE, SF, and SFE models is described below.

\subsection{The skew errors (SE) model}
\label{s:SE}

\noindent
The SE model admits a straightforward hierarchical representation:
\begin{eqnarray}
\bY_{ij} \mid \bU_{ij}, w_{ij}		
		&\sim&	N_p\left(\bmu_i + \bDelta_i |\bU_{ij}|, \frac{1}{w_{ij}} \bSigma_i\right), \nonumber\\
\bU_{ij} \mid w_{ij}	
		&\sim&	N_r \left(\bzero, \frac{1}{w_{ij}} \bI_r\right), \nonumber\\
w_{ij}	&\sim&	\mbox{gamma}\left(\frac{\nu_i}{2}, \frac{\nu_i}{2}\right),
\label{SE}
\end{eqnarray}
where $\bSigma_i = \bB_i\bB_i^T + \bD_i$. 
\newline

\subsubsection{Cycle One}

In the first cycle, the missing data are $Z_{ij}$, $\bU_{ij}$, and $w_{ij}$. This is essentially identical to a traditional FM-CFUST model. Hence from \citet{J164}, the conditional expectations required for the E-step are given by
\begin{eqnarray}
z_{ij}^{(k)} &=&	E_{\Psi^{(k)}} \left[z_{ij}=1 \mid \by_j\right] 
			= \frac{\pi_i^{(k)} f_{\mbox{\tiny{CFUST}}_{p,r}}
			(\by_j; \bmu_i^{(k)}, \bSigma_i^{(k)}, \bDelta_i^{(k)}, \nu_i^{(k)})}
			{\sum_{i=1}^g \pi_i^{(k)} f_{\mbox{\tiny{CFUST}}_{p,r}}
			(\by_j; \bmu_i^{(k)}, \bSigma_i^{(k)}, \bDelta_i^{(k)}, \nu_i^{(k)})}, 
			\label{e0}\\
w_{ij}^{(k)}	&=&	E_{\Psi^{(k)}} \left[w_{ij} \mid \by_j, z_{ij}=1\right] 
			\nonumber\\&&
			= \left(\frac{\nu_i^{(k)} + p}{\nu_i^{(k)} + d_{ij}^{(k)}}\right) 
			\frac{T_r\left(\bq_{ij}^{(k)} \sqrt{\frac{\nu_i^{(k)}+p+2}{\nu_i+d_{ij}^{(k)}}}; 
				\bzero, \bLambda_i^{(k)}, \nu_i^{(k)}+p+2\right)}
				{T_r\left(\bq_{ij}^{(k)} \sqrt{\frac{\nu_i^{(k)}+p}{\nu_i+d_{ij}^{(k)}}}; 
				\bzero, \bLambda_i^{(k)}, \nu_i^{(k)}+p\right)}, 
				\label{e1}\\
\bu_{ij}^{(k)}	&=&	E_{\Psi^{(k)}} \left[w_{ij} \bU_{ij}
		\mid \by_j, z_{ij}=1 \right]
			= w_{ij}^{(k)} E\left[ \ba_{ij}^{(k)}\right], 	\label{e4}\\
\bu_{ij}^{*^{(k)}} &=&	E_{\Psi^{(k)}} \left[w_{ij} \bU_{ij} \bU_{ij}^T
		\mid, \by_j, z_{ij}=1 \right]
			= w_{ij}^{(k)} E\left[ \ba_{ij}^{(k)} \ba_{ij}^{(k)^T} \right],	\label{e5}
\end{eqnarray} 
where
\begin{eqnarray}
d_{ij}^{(k)}	&=&	(\by_j - \bmu_i^{(k)})^T \bOmega_i^{(k)^{-1}} (\by_i - \bmu_i^{(k)}), 
		\nonumber\\
\bq_{ij}^{(k)}	&=& \bDelta_i^{(k)^T} \bOmega_i^{(k)^{-1}} (\by_j-\bmu_i^{(k)}), 
		\nonumber\\
\bLambda_i^{(k)}	&=& \bI_r - \bDelta_i^{(k)^T} \bOmega_i^{(k)^{-1}} \bDelta_i^{(k)}, 
		\nonumber\\
\bOmega_i^{(k)}	&=& \bSigma_i^{(k)} + \bDelta_i^{(k)} \bDelta_i^{(k)^T}, 
		\nonumber
\end{eqnarray}
and 
\begin{eqnarray}
\ba_{ij}^{(k)}	&\sim& tt_r\left(\bq_{ij}^{(k)}, 
		\left(\frac{\nu_i^{(k)} + d_{ij}^{(k)}}{\nu_i^{(k)}+p+2}\right) \bLambda_i^{(k)}, 
		\nu_i^{(k)}+p+2; \mathbb{R}^+\right). 
\end{eqnarray}
In the above, $f_{\mbox{\tiny{CFUST}}}(\cdot)$ denotes the density of a CFUST distribution, $T_r(\cdot)$ denotes the distribution function of an $r$-dimensional $t$-distribution, and $tt_r(\cdot; \mathbb{R}^+)$ denotes the $r$-dimensional truncated $t$-density truncated to the positive hyperplane.   
\newline

\noindent
The M-step in this cycle is the same as in the case of the traditional FM-CFUST model, except that the update of the scale matrix $\bSigma_i$ is not used (but still needs to be calculated as it is required for the M-step in the second cycle). It follows that the M-step is given by
\begin{eqnarray}
\pi_i^{(k+1)}	&=&	\frac{1}{n} \sum_{j=1}^n z_{ij}^{(k)}, \nonumber\\
\bmu_i^{(k+1)}	&=&	\frac{\sum_{j=1}^n z_{ij}^{(k)} w_{ij}^{(k)}\by_j 
		- \bDelta_i^{(k)} \sum_{j=1}^n z_{ij}^{(k)} \bu_{ij}^{(k)}} 
		{\sum_{j=1^n}^n z_{ij}^{(k)} w_{ij}^{(k)}}, \nonumber\\
\bDelta_i^{(k+1)}	&=&	\left[\sum_{j=1}^n z_{ij}^{(k)} 
		(\by_i - \bmu_i^{(k)}) \bu_{ij}^{(k)^T}\right]	
		\left[\sum_{j=1}^n z_{ij}^{(k)} \bu_{ij}^{*^{(k)}}\right]^{-1}. \nonumber
\end{eqnarray}
\newline

\noindent
An update of the degrees of freedom $v_i$ is obtained by solving the following equation. 
\begin{eqnarray}
0 &=& \left(\sum_{i=1}^n z_{ij}^{(k)}\right) 
		\left[\log\left(\frac{\nu_i}{2}\right) 
		- \psi\left(\frac{\nu_i}{2}\right) + 1\right] \nonumber\\ &&
		+ \sum_{j=1}^n \tau_{ij}^{(k)} \left[\psi\left(\frac{\nu_i^{(k)}+p}{2}\right)
		- \log\left(\frac{\nu_i^{(k)}+\eta_{ij}^{(k)}}{2}\right)
		- \left(\frac{\nu_i^{(k)}+p}{\nu_i^{(k)}+\eta_{ij}^{(k)}}\right) \right],	\nonumber
\end{eqnarray}
where
\begin{eqnarray}
\eta_{ij}^{(k)} &=& \left(\by_j - \bmu_i^{(k)}\right)^T 
\left(\bB_i^{(k)} \bOmega_i^{(k)} \bB_i^{(k)^T} + \bD_i^{(k)}\right)^{-1} 
\left(\by_j - \bmu_i^{(k+1)}\right), \nonumber\\
\bOmega_i^{(k)}	&=& \bI_q + \bDelta_i^{(k)} \bDelta_i^{(k)^T}, \nonumber
\end{eqnarray}
and where $\psi(\cdot)$ is the digamma function. 
\newline
\newline

\noindent
Although not explicitly used in the AECM algorithm, the update for the scale matrix $\bSigma_i$ is used implicitly in the M-step of the second cycle and is given by 
\begin{eqnarray}
\bSigma_i^{(k+1)} &=& \frac{\sum_{j=1}^n z_{ij}^{(k)} 
		\left[(\by_i - \bmu_i^{(k+1)}) (\by_j - \bmu_i^{(k+1)})^T 
		- \bDelta_i^{(k)} \bu_{ij}^{(k)^T} \bDelta_i^{(k)^T}\right]}
		{\sum_{j=1}^n z_{ij}^{(k)}}. 
\nonumber
\end{eqnarray}

\subsubsection{Cycle Two}

In the second cycle, the missing data are those in the first cycle and also the latent factors; that is, they include $Z_{ij}$, $\bU_{ij}$, $w_{ij}$, and $\bX_{ij}$. In this cycle, we obtain updated estimates for the parameters $\bB_i$ and $\bD_i$. 
These are analogous to those in the case of the MFA model and are given, respectively, by
\begin{eqnarray}
\bB_i^{(k+1)} &=& \bSigma_i^{(k+1)} \bbeta_i^T \bA_i^{-1}, \nonumber\\
\bD_i^{(k+1)} &=& \mbox{diag}\left(\bSigma_i^{(k+1)} 
		- \bB_i^{(k+1)} \bbeta_i \bSigma_i^{(k+1)}\right),
\nonumber
\end{eqnarray}   
where
\begin{eqnarray}
\bbeta_i &=& \bB_i^{(k+1)} \left(\bB_i^{(k+1)}\bB_i^{(k+1)^T} + \bD_i^{(k)}\right)^{-1}, 
	\nonumber\\
\bA_i &=& \bI_p - \bbeta_i\bB_i^{(k+1)} + \bbeta_i\bSigma_i^{(k+1)} \bbeta_i^T.
\end{eqnarray}
\newline

\subsection{The skew factors (SF) model}
\label{s:SF}

Not surprisingly, the expressions of the conditional expectations and the updated estimate of parameters on the E- and M-steps of the AECM algorithm for the SE model are not as straightforward as for the SF model. The technical details can be found in \citet{J638}. In brief, we exploit the hierarchical representation given by
\begin{eqnarray}
\bY_j	\mid \bx_{ij}, w_{ij}, Z_{ij} =1	&\sim& 
	N_p\left(\bB\bx_{ij} + \bmu_i, \frac{1}{w_{ij}} \bD_i\right),
	\nonumber\\
\bX_{ij}	\mid \bu_{ij}, w_{ij}, Z_{ij} =1	&\sim& 
	N_q\left(\bDelta_i|\bu_{ij}|, \frac{1}{w_{ij}} \bI_q\right),
	\nonumber\\
\bU_{ij} \mid w_{ij}, Z_{ij} =1	&\sim&	
	N_r\left(\bI_r, \frac{1}{w_{ij}} \bI_r\right),
	\nonumber\\
W_{ij} \mid Z_{ij} =1	&\sim&	\mbox{gamma}\left(\frac{\nu_i}{2}, \frac{\nu_i}{2}\right),
		\nonumber\\
Z_{ij} =1 &\sim& \mbox{Multi}_g(1; \bpi). 
\label{hierarchical}
\end{eqnarray}  

It follows that the E-step involves three extra conditional expectations compared to the SE model. Thus, we need to compute (\ref{e0}) to (\ref{e5}), but with $\bDelta_i$ replaced by $\bDelta_i^* = \bB_i\bDelta_i$. Note that this implies corresponding changes to $q_{ij}$, $\bLambda_i$, and $\bOmega_i$. The three additional conditional expectations are due to the latent factors and are given by 
$\bx_{ij}^{(k)} = E_{\Psi^{(k)}} \left[w_{ij} \bX_{ij} \mid, \by_j, z_{ij}=1 \right]$, 
$\tilde{\bx}_{ij}^{(k)} = E_{\Psi^{(k)}} \left[w_{ij} \bX_{ij} \bU_{ij}^T
		\mid, \by_j, z_{ij}=1 \right]$, and
$\bx_{ij}^{*^{(k)}} = E_{\Psi^{(k)}} \left[w_{ij} \bU_{ij} \bU_{ij}^T
		\mid, \by_j, z_{ij}=1 \right]$. 	
It can be shown that 
\begin{eqnarray}
\bx_{ij}^{(k)}	&=&	w_{ij}^{(k)} \bC_{i}^{(k)} \bB_{i}^{(k)^T} \bD_{i}^{(k)^{-1}} 
		\left(\by_j - \bmu_{i}^{(k)}\right) 
		+ \bC_{i}^{(k)} \bDelta_{i}^{(k)} \bu_{ij}^{(k)}, 
		\label{X}\\
\tilde{\bx}_{ij}^{(k)}	&=& \bC_{i}^{(k)} \bB_{i}^{(k)^T} \bD_{i}^{(k)^{-1}} 
		\left(\by_j - \bmu_{i}^{(k)}\right) \bu_{ij}^{(k)^T} 
		+ \bC_{i}^{(k)} \bDelta_{i}^{(k)} \bu_{ij}^{*^{(k)}},
		\label{Xt}\\
\bx_{ij}^{*^{(k)}}	&=& \bx_{ij}^{(k)} \left(\by_j - \bmu_{i}^{(k)}\right)^T 
		\bB_{i}^{(k)} \bC_{i}^{(k)^T} 
		+ \tilde{\bx}_{ij}^{(k)} \bDelta_{i}^{(k)^T}  
		\bC_{i}^{(k)^T} + \bC_{i}^{(k)}, 
		\label{Xs}
\end{eqnarray}
where $\bC_i^{(k)^{-1}} = \bB_i^{(k)^T} \bD_i^{(k)^{-1}} \bB_i^{(k)} + \bI_q$. 
\newline
	
\noindent
For the M-step, the expression for the updated estimate of the parameters are quite similar to the SE model and are given by
\begin{eqnarray}
\pi_i^{(k+1)}	&=&	\frac{1}{n} \sum_{j=1}^n z_{ij}^{(k)}. \label{pi}\\
\bDelta_i^{(k+1)}	&=&	\left[\sum_{j=1}^n z_{ij}^{(k)} \tilde{\bx}_{ij}^{(k)}\right]	
		\left[\sum_{j=1}^n z_{ij}^{(k)} \bu_{ij}^{*^{(k)}}\right]^{-1}. \nonumber\\
\bB_i^{(k+1)}	&=&	\left[\sum_{j=1}^n z_{ij}^{(k)} 
		\left(\by_j-\bmu_i^{(k+1)}\right) \bx_{ij}^{(k)^T}\right]	
		\left[\sum_{j=1}^n z_{ij}^{(k)} \bx_{ij}^{*^{(k)}}\right]^{-1}. \nonumber\\
\bmu_i^{(k+1)}	&=&	\frac{\sum_{j=1}^n z_{ij}^{(k)} w_{ij}^{(k)}\by_j 
		- \bB_i^{(k)}\bDelta_i^{(k)} \sum_{j=1}^n z_{ij}^{(k)} \bu_{ij}^{(k)}} 
		{\sum_{j=1}^n z_{ij}^{(k)} w_{ij}^{(k)}}. \label{mu}\\
\bD_i^{(k+1)} &=& \mbox{diag}\left(\bd_i^{(k+1)}\right), \nonumber
\end{eqnarray} 
where
\begin{eqnarray}
\bd_i^{(k+1)}	&=&	\mbox{diag} \left\{
		\sum_{j=1}^n z_{ij}^{(k)} \left[ w_{ij}^{(k)} 
		\left(\by_j - \bmu_i^{(k+1)}\right) \left(\by_j-\bmu_i^{(k+1)}\right)^T
		- \bB_i^{(k)} \bx_{ij}^{(k)} \left(\by_j-\bmu_i^{(k+1)}\right)^T 
		\right.\right. \nonumber\\ && \left.\left.
		- \left(\by_j-\bmu_i^{(k+1)}\right) \bx_{ij}^{(k)^T} \bB_i^{(k)^T} 
		- \bB_i^{(k)} \bx_{ij}^{*^{(k)}} \bB_i^{(k)} \right] \right\}
		\left[\sum_{j=1}^n z_{ij}^{(k)}\right]^{-1}. \label{nu}
\end{eqnarray} 
Concerning the update for the degrees of freedom, it is the same as for the SE model.
\newline

\subsection{The skew factors and errors (SFE) model}
\label{s:SFE}

The SFE model is a combination of the SE and SF models. 
It follows from (\ref{error_factor}) that it can be expressed in a slightly more complicated  hierarchical form than (\ref{hierarchical}). An extra level is required for the skewing variable $\bU_{1ij}$ for the errors. It follows that 
\begin{eqnarray}
\bY_j	\mid \bx_{ij}, w_{ij}, Z_{ij} =1	&\sim& 
	N_p\left(\bmu_i + \bB\bx_{ij} + \bDelta_{1i}|\bu_{1ij}|, \frac{1}{w_{ij}} \bD_i\right),
	\nonumber\\
\bX_{ij}	\mid \bu_{0ij}, w_{ij}, Z_{ij} =1	&\sim& 
	N_q\left(\bDelta_{0i}|\bu_{0ij}|, \frac{1}{w_{ij}} \bI_q\right),
	\nonumber\\
\bU_{0ij} \mid w_{ij}, Z_{ij} =1	&\sim&	
	N_r\left(\bI_r, \frac{1}{w_{ij}} \bI_r\right), 
	\nonumber\\
\bU_{1ij} \mid w_{ij}, Z_{ij} =1	&\sim&	
	N_s\left(\bI_s, \frac{1}{w_{ij}} \bI_s\right), 
	\nonumber\\
W_{ij} \mid Z_{ij} =1	&\sim&	\mbox{gamma}\left(\frac{\nu_i}{2}, \frac{\nu_i}{2}\right),
		\nonumber\\
Z_{ij} =1 &\sim& \mbox{Multi}_g(1; \bpi). 
\label{SFE_h}
\end{eqnarray}  

According to the above specification, although $\bu_{0ij}$ and $\bu_{1ij}$ are uncorrelated, they are not independent. 
Due to this, the calculation of the conditional expectation of $\bu_{0ij}$ and of $\bu_{1ij}$ is performed jointly and thus involves evaluating $(r+s)$-dimensional integrals. 

In the first cycle of the AECM algorithm for the SFE model, we proceed in a similar manner as for the SF model. However, $\bDelta_i^*$ now involves both $\bDelta_{0i}$ and $\bDelta_{1i}$, that is, it is a $p \times (r+s)$ matrix given by $\bDelta_i^* = [\bB_i\bDelta_{0i} \;\; \bDelta_{1i}]$.  
We also let $\bSigma_i^* = \bB_i \bB_i^T + \bD_i^T$. 
In a similar way, $\bOmega_i^*$, $\bq_i^*$, $d_{ij}^{*^(k)}$, and $\bLambda_i^*$ are defined in terms of $\bSigma_i^*$ and $\bDelta_i^*$ (in place of the usual $\bSigma_i$ and $\bDelta_i$, respectively). 
Thus, on the $k$th iteration of the E-step, the following conditional expectations are required: 
\begin{eqnarray}
z_{ij}^{(k)} &=&	E_{\Psi^{(k)}} \left[z_{ij}=1 \mid \by_j\right] 
			= \frac{\pi_i^{(k)} f_{\mbox{\tiny{CFUST}}_{p,r}}
			(\by_j; \bmu_i^{(k)}, \bSigma_i^{*^(k)}, \bDelta_i^{*^(k)}, \nu_i^{(k)})}
			{\sum_{i=1}^g \pi_i^{(k)} f_{\mbox{\tiny{CFUST}}_{p,r}}
			(\by_j; \bmu_i^{(k)}, \bSigma_i^{*^(k)}, \bDelta_i^{*^(k)}, \nu_i^{(k)})}, 
	\label{se0}\\
w_{ij}^{(k)}	&=&	E_{\Psi^{(k)}} \left[w_{ij} \mid \by_j, z_{ij}=1\right] \nonumber\\		
			&=& \left(\frac{\nu_i^{(k)} + p}{\nu_i^{(k)} + d_{ij}^{*^(k)}}\right) 
			\frac{T_r\left(\bq_{ij}^{*^(k)} \sqrt{\frac{\nu_i^{(k)}+p+2}{\nu_i+d_{ij}^{(k)}}}; 
				\bzero, \bLambda_i^{*^(k)}, \nu_i^{(k)}+p+2\right)}
				{T_r\left(\bq_{ij}^{*^(k)} \sqrt{\frac{\nu_i^{(k)}+p}{\nu_i+d_{ij}^{(k)}}}; 
				\bzero, \bLambda_i^{*^(k)}, \nu_i^{(k)}+p\right)}, 
		\label{se1}\\
\bu_{ij}^{(k)}	&=&	E_{\Psi^{(k)}} \left[w_{ij} \bU_{ij}
		\mid \by_j, z_{ij}=1 \right]
			= w_{ij}^{(k)} E\left[ \ba_{ij}^{(k)}\right], 	
	\label{se4}\\
\bu_{ij}^{*^{(k)}} &=&	E_{\Psi^{(k)}} \left[w_{ij} \bU_{ij} \bU_{ij}^T
		\mid, \by_j, z_{ij}=1 \right]
			= w_{ij}^{(k)} E\left[ \ba_{ij}^{(k)} \ba_{ij}^{(k)^T} \right], 
	\label{se5}
\label{SFE-E}
\end{eqnarray} 
where
\begin{eqnarray}
\ba_{ij}^{(k)}	&\sim& tt_{r+s}\left(\bq_{ij}^{*^(k)}, 
		\left(\frac{\nu_i^{(k)} + d_{ij}^{*^(k)}}{\nu_i^{(k)}+p+2}\right) \bLambda_i^{(k)}, 
		\nu_i^{(k)}+p+2; \mathbb{R}^+\right). 
\end{eqnarray}

The required conditional expectations related to $\bu_{0ij}$ and $\bu_{1ij}$ are extracted from (\ref{se4}) and (\ref{se5}) above using
\begin{eqnarray}
\bu_{ij}^{(k)}	&=&	
		\left[\begin{array}{c}\bu_{0ij}^{(k)}\\\bu_{1ij}^{(k)}\end{array}\right],
\label{se6} \\
\bu_{ij}^{*^(k)}	&=&	
		\left[\begin{array}{cc}
				\bu_{0ij}^{*^(k)} & \bu_{3ij}^{(k)} \\
				\bu_{3ij}^{(k)^T}  & \bu_{1ij}^{*^(k)}
		\end{array}\right].
\label{se7}
\end{eqnarray}

For the first cycle of the AECM algorithm, the M-step proceeds in a similar way to the SF model described in Section \ref{s:SF}. The updated estimates for $\pi_i$, $\bmu_i$, and $\nu_i$ are calculated using (\ref{pi}), (\ref{mu}), and (\ref{nu}), respectively, but with $\bDelta_i^{(k)}$ replaced by $\bDelta_i^{*^{(k)}}$.   

In the second cycle, we calculate the conditional expectations related to the factors $\bX_{ij}$ and compute the updated estimate for $\bB_i$, $\bD_i$, $\bDelta_{0i}$, and $\bDelta_{1i}$. The four conditional expectations required on the E-step are analogous to (\ref{X}), (\ref{Xs}), and with (\ref{Xt}) separated into $\tilde{\bX}_{0ij}$ and $\tilde{\bX}_{1ij}$. It can be shown that they are given by
\begin{eqnarray}
\bx_{ij}^{(k)}	&=&	w_{ij}^{(k)} \bC_{i}^{(k)} \bB_{i}^{(k)^T} \bD_{i}^{(k)^{-1}} 
		\left(\by_j - \bmu_{i}^{(k)}\right) 
		+ \bC_{i}^{(k)} \bB_{i}^{(k)^T} \bD_{i}^{(k)^{-1}} \bDelta_{1i}^{(k)} \bu_{1ij}^{(k)} 
		+ \bC_{i}^{(k)} \bDelta_{0i}^{(k)} \bu_{2ij}^{(k)}, \nonumber\\
\tilde{\bx}_{0ij}^{(k)}	&=& \bC_{i}^{(k)} \bB_{i}^{(k)^T} \bD_{i}^{(k)^{-1}} 
		\left(\by_j - \bmu_{i}^{(k)}\right) \bu_{0ij}^{(k)^T} 
		- \bC_{i}^{(k)} \bB_{i}^{(k)^T} \bD_{i}^{(k)^{-1}} \bDelta_{1i}^{(k)} \bu_{1ij}^{(k)}
		+ \bC_{i}^{(k)} \bDelta_{0i}^{(k)} \bu_{0ij}^{*^{(k)}}, \nonumber\\
\tilde{\bx}_{1ij}^{(k)}	&=& \bC_{i}^{(k)} \bB_{i}^{(k)^T} \bD_{i}^{(k)^{-1}} 
		\left(\by_j - \bmu_{i}^{(k)}\right) \bu_{1ij}^{(k)^T} 
		- \bC_{i}^{(k)} \bB_{i}^{(k)^T} \bD_{i}^{(k)^{-1}} \bDelta_{1i}^{(k)} \bu_{1ij}^{*^(k)}
		+ \bC_{i}^{(k)} \bDelta_{0i}^{(k)} \bu_{0ij}^{*^{(k)}}, \nonumber\\
\bx_{ij}^{*^{(k)}}	&=& \bx_{ij}^{(k)} \left(\by_j - \bmu_{i}^{(k)}\right)^T 
		\bD_i^{(k)^{-1}} \bB_{i}^{(k)} \bC_{i}^{(k)^T} 
		- \tilde{\bx}_{1ij}^{(k)} \bDelta_{1i}^{(k)^T} 
		  \bD_i^{(k)^{-1}}  \bB_{i}^{(k)} \bC_{i}^{(k)^T} 
		\nonumber\\ &&
		+ \; \tilde{\bx}_{0ij}^{(k)} \bDelta_{0i}^{(k)^T}  
		\bC_{i}^{(k)^T} 
		+ \bC_{i}^{(k)}, \nonumber
\end{eqnarray}
where $\bC_i^{(k)}$ is the same as for the SF model.

\section{Conclusions}
\label{sec:con}

In this paper, we described and discussed the differences between placing the assumption of skewness on the factors (SF) or/and the errors (SE) in mixtures of skew factor analyzers. In doing so, we introduced the more general skew factor and error (SFE) MFA approach where both the factors and the errors have a skew component distribution. 
Parameter estimation via an EM-type algorithm for these approaches was discussed and an AECM algorithm was derived for the SFE model. 
The implementation of the EM algorithm was easier to undertake for the SE model than for the SF and SFE models.  
We note that given the same values of $g$, $p$, $q$, and $r$, the SE model has a higher number of free parameters compared to the SF model. 
The practical implications of these formulations will be treated in a forthcoming manuscript, based on simulations and real data applications.


\end{document}